\documentclass{iopjournal}
\usepackage{subfig}
\usepackage{graphicx}
\usepackage{eurosym}
\usepackage{amsmath}
\usepackage{siunitx}
\usepackage{multirow}

\begin{document}

\articletype{Paper} %	 e.g. Paper, Letter, Topical Review...

\title{CRYSP: a Total-Body PET based on cryogenic
cesium iodide crystals}

\author{S.R. Soleti$^{1,2,*}$\orcid{0000-0002-5526-1414},  P. Dietz$^{1}$\orcid{0009-0002-1803-0892}, R. Esteve$^{3}$, J. García-Barrena$^{3}$, V. Herrero$^{3}$, F. Lopez$^{1}$, F.~Monrabal$^{1,2}$\orcid{0000-0002-4047-5620}, L.~Navarro-Cozcolluela$^{1,4}$, E.~Oblak$^{1}$, J. Pelegrín$^{1}$\orcid{0000-0002-7589-5940}, J. Renner$^{5}$\orcid{0000-0003-1843-2015}, J.~Toledo$^{3}$, S.~Torelli$^{1}$\orcid{0000-0003-3622-3524}, and J.J.~G\'omez-Cadenas$^{1,2}$\orcid{0000-0002-8224-7714}}

\affil{$^1$Donostia International Physics Center, San Sebasti\'an / Donostia, E-20018, Spain}

\affil{$^2$Ikerbasque (Basque Foundation for Science), Bilbao, E-48009, Spain}

\affil{$^3$Instituto de Instrumentaci\'on para Imagen Molecular (I3M), Centro Mixto CSIC - Universitat Polit\`ecnica de Val\`encia, Valencia, E-46022, Spain}

\affil{$^4$Universidad del País Vasco (UPV/EHU), San Sebasti\'an / Donostia, E-20018, Spain}

\affil{$^5$Instituto de F\'isica Corpuscular (IFIC), CSIC \& Universitat de Val\`encia, Paterna, E-46980, Spain}

\affil{$^*$Author to whom any correspondence should be addressed.}

\email{roberto.soleti@dipc.org}


\begin{abstract}
\textit{Objective.} Total Body PET (TBPET) scanners have the potential to substantially reduce both acquisition time and administered radiation dose, owing to their high sensitivity. However, their widespread clinical adoption is hindered by the high cost of currently available systems. This work explores the use of pure cesium iodide (CsI) monolithic crystals operated at cryogenic temperatures as a cost-effective alternative to rare-earth scintillators for TBPET. \textit{Approach.} We investigate the performance of pure CsI crystals operated at cryogenic temperatures ($\sim$100 K), where they achieve a light yield of approximately $10^5$ photons/MeV. The implications for energy resolution, spatial resolution (including depth-of-interaction capability), and timing performance are assessed, with a view toward their integration into a TBPET system. \textit{Main results.} Cryogenic CsI crystals demonstrated energy resolution below 7\% and coincidence time resolution at the nanosecond level, despite their relatively slow scintillation decay time. A Monte Carlo simulation of monolithic CsI crystals shows that a millimeter-scale spatial resolution in all three dimensions can be obtained. These characteristics indicate that high-performance PET imaging is achievable with this technology. \textit{Significance.} A TBPET scanner based on cryogenic CsI monolithic crystals could combine excellent imaging performance with significantly reduced detector costs, enabling broader accessibility and accelerating the adoption of TBPET in both clinical and research settings.
\end{abstract}

\section{Introduction}
Positron emission tomography (PET) is a powerful imaging technique central to medical diagnostic applications, employing positron-emitting radionuclides to track biologically active molecules. The annihilation of the positrons produced by the radiotracer with the electrons of the surrounding environment generates back-to-back gamma rays of 511~keV, which are detected to create detailed images of functional processes within the body.

\begin{figure}[htbp]
\centering
\includegraphics[width=0.65\linewidth]{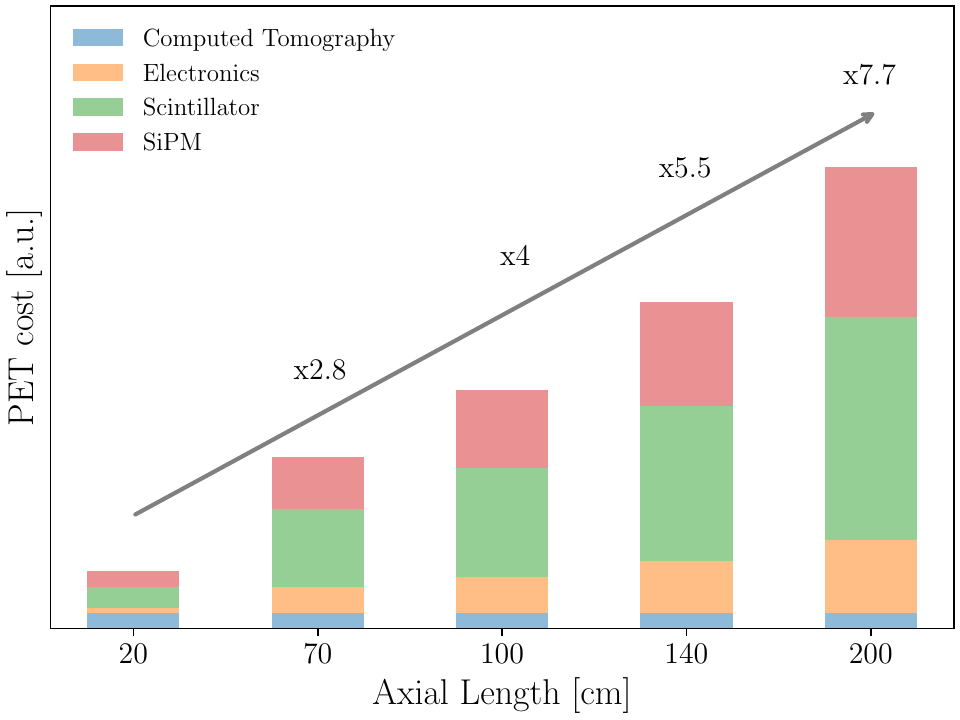}
\caption{Increase in component costs for a 70-cm, 100-cm, 140-cm and 200-cm system versus a system with 20 cm axial length; the y-axis is the system cost in relative units. Data taken from ref.~\cite{vandenberghe2020state}.}\label{fig:cost}
\end{figure}

The last decade has witnessed the development of Total Body PET (TBPET) scanners with axial fields-of-view (AFOV) between 3 and 7 times that of conventional PET (CPET) apparatus. Such devices detect a large fraction of the emitted photons, thus dramatically increasing sensitivity, and allowing for simultaneous dynamic acquisition from all tissues of interest~\cite{vandenberghe2020state, nadig2022hybrid}. TBPETs also offer the flexibility of optimizing signal-to-noise ratio, reducing the acquisition time, or decreasing the amount of administered radiation dose.

Prime examples of such TBPETs are the United Imaging {uEXPLORER}~\cite{cherry2018total}, and Siemens {Biograph Vision Quadra}~\cite{prenosil2022performance}. The AFOV of uEXPLORER is 194 cm, while the AFOV of the smaller Quadra is 106~cm, to be compared with an AFOV of 20-30 cm typical of a CPET (e.g., the new generation General Electric Omni Legend 32~\cite{yamagishi2023performance} has an AFOV of 32~cm). Both TBPET systems use arrays of pixelated lutetium-yttrium oxyorthosilicate (LYSO) crystals, read out by SiPMs and custom electronics, with very high sensitivity and excellent time-of-flight (TOF) resolution. The approximate increase in cost of the uEXPLORER (Quadra) compared with a typical CPET is roughly 8 (4) times (see  fig.~\ref{fig:cost}), well above what can be afforded by most hospitals.

 \begin{figure}[htbp]
 \centering
 \includegraphics[width=0.65\linewidth]{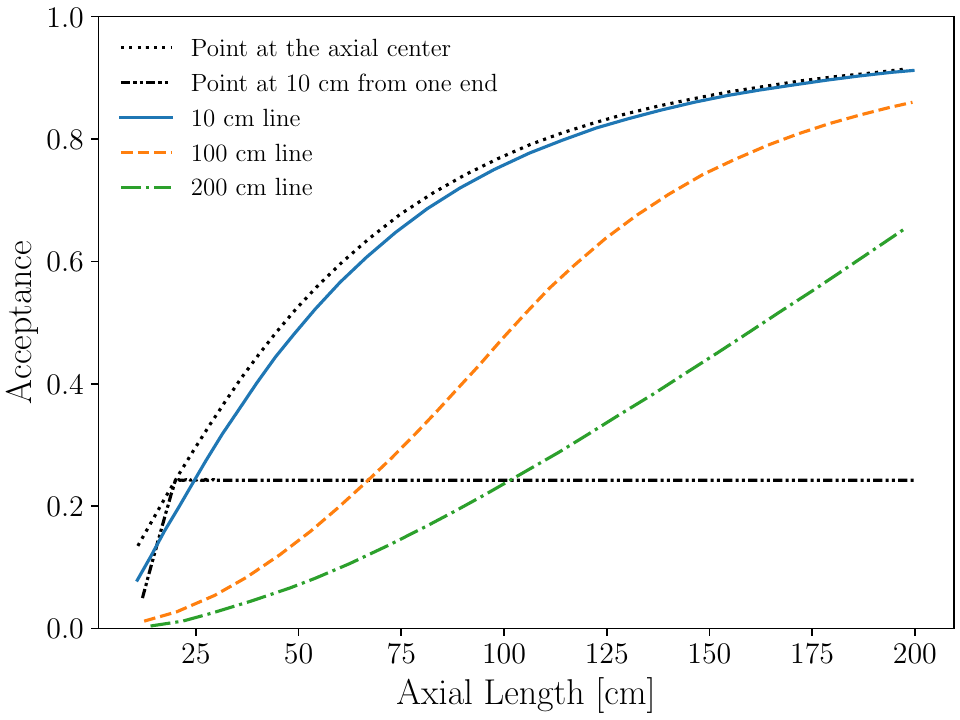}
 \caption{The geometrical acceptance for a point-like source placed at the axial center, a point-like source placed 10~cm from one end of the scanner, and line sources of 10 cm, 100 cm and 200 cm length, as a function of the axial length. It is calculated in the transverse center of a PET scanner with a diameter of 80 cm.}\label{fig:geomacc}
 \end{figure}

The bulk of the cost in both systems comes from the scintillator, followed by the cost of the SiPMs and electronics. Of the three items, the cost of SiPMs, in large quantities, is  roughly proportional to the total surface of chips deployed (not to the size of the chip), and thus proportional to the size and coverage of the scanner. However, the cost of SiPMs decreases about 5-10 \% annually, with current prices about one third of those ten years ago. Thus their relative contribution to the total cost of future TBPETs will steadily decrease with time. Instead, the cost of crystals and that of the readout electronics has remained stable, and in the case of LYSO, significantly expensive. Thus, the first priority of any putative low-cost TBPET is to find significantly cheaper alternatives to LYSO. Notice that, in addition to the high cost, LYSO is a rare earth whose production and manufacturing is heavily concentrated. Last but not least, the cost of the readout electronics can be decreased for large scale production developing specific ASICs, and, importantly, reducing the number of electronics channels.  

Naturally, one also desires performance comparable to that of current PET scanners. This, in turn, requires quantifying the contribution of the extended axial length of a TBPET scanner to the quality of image reconstruction.

{Figure~\ref{fig:geomacc} shows the geometrical acceptance for two point-like sources (serving as rough approximations of small organs), one positioned at the center of a scanner with an 80~cm diameter and the other located 10~cm from one end, as a function of the axial length. For the centrally placed source, approximately 80\% of the solid angle is already covered at a scanner length of 100~cm. In contrast, for the source near the edge (e.g., the brain in a TBPET), the acceptance quickly plateaus: photons emitted toward the open end of the scanner escape detection, and increasing the AFOV beyond this point offers no additional benefit. Overall, most of the gain in geometrical acceptance is achieved within the first 100~cm of axial coverage, with only marginal improvements beyond that length.}

The choice for a PET scanner of 100~cm length is also motivated by the approximate axial distribution of whole-body organs of interest (e.g., brain to pelvis) for typical human height. The figure shows that, with an extended source of 100 cm, one gains in geometrical acceptance from 50\% (100 cm scanner), to about 80\% (200 cm scanner), a modest increased compared with cost. Thus, an AFOV of about 100~cm as implemented by the Quadra appears as a good compromise between cost and performance and a reasonable target for any TBPET proposal.

 \begin{figure*}[htbp]
    \centering
    \subfloat[]{\includegraphics[width=0.34\linewidth]{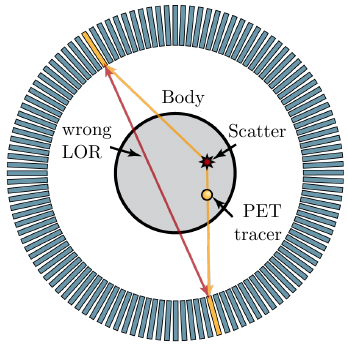}}
    \hfil
    \subfloat[]{\includegraphics[width=0.45\linewidth]{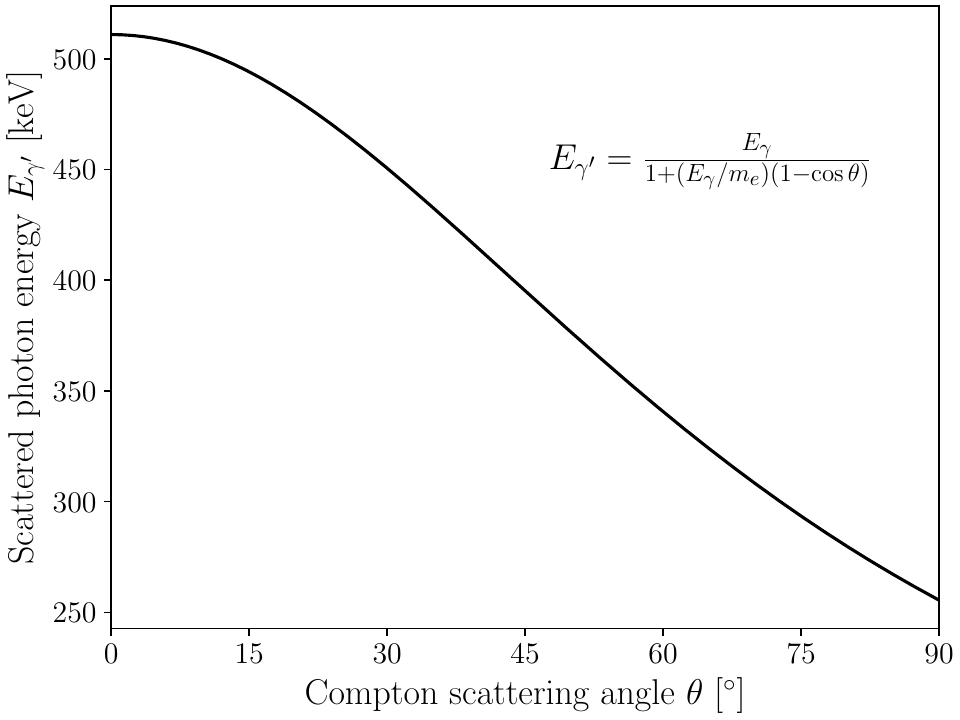}}
    \caption{(a) Compton scatter detection in PET. In Compton scattering a photon (in yellow) undergoes an interaction within the patient’s body where it changes direction and loses energy. Thus, the PET scanner reconstructs a wrong LOR (in red), affecting the signal-to-background ratio of the reconstructed image. (b) The residual energy $E_{\gamma^{\prime}}$ of a 511-keV incoming photon after Compton scattering with a given angle $\theta$.}\label{fig:mscat}
 \end{figure*}

 %  \begin{figure}[htbp]
 % \centering
 % \includegraphics[width=0.6\linewidth]{images/parallax.pdf}
 % \caption{Mean parallax error $\Delta\theta$, defined as the angular distance between the ideal LOR and the reconstructed LOR, as a function of the LOR axial angle. The result was obtained by simulating a PET scanner with an axial length of 1~m and a diameter of 0.8 m, made of $3\times3\times22.4$~mm$^{3}$ LYSO pixels.}\label{fig:pe}
 % \end{figure}

But is that extra length worth the added cost? Addressing this question requires consideration not only of the improved geometrical acceptance but also of the accompanying increase in gamma-ray absorption and Compton scattering (CS) within the patient's body, as well as the effects of parallax error (PE). These effects become more pronounced as gammas travel at increasingly oblique angles. Increased gamma absorption results in loss of events and is an unavoidable effect. CS and PE introduce an error in the reconstruction of the line-of-response (LOR), and thus contribute to image blurring. Both effects can be minimized by instrumental choices. 

The impact of CS, exemplified in fig. \ref{fig:mscat}, can be minimized by choosing a narrow window around 511~keV, which selects events that deposit all their energy in the crystal. The efficiency of this selection, in turn, improves with the energy resolution of the crystal. Conversely, if one selects events in a fixed energy window, the number of events which have suffered scattering in the patient's body decreases as energy resolution improves.  

Conventional PET scanners are made of pixelated detectors, with crystals of small transverse dimensions and a thickness of typically 2 radiation lengths (e.g., $3.2 \times 3.2 \times 20$~mm$^3$ for the Quadra). For a gamma impinging at angles normal to the crystal surface (small transaxial angles), the resolution in the transaxial coordinates depends only on the transverse size of the pixel, and the resolution on the depth-of-interaction (d.o.i.), {which depends on the pixel thickness, does not significantly affect the reconstruction.}
% , Specifically, $\sigma_{x,z} =  s/\sqrt{12}$~mm and  $\sigma_{r} =  t/\sqrt{12}$~mm. For example, in the case of Quadra, one obtains, $\sigma_{x,z} =  3.2/\sqrt{12} = 0.9$~mm and  $\sigma_{r} =  t/\sqrt{12}=5.8$~mm. 
While d.o.i. information can be incorporated into pixelated  detectors through dedicated schemes (such as dual-ended readout, phoswich configurations, or depth-segmented crystals), the standard pixelated blocks used in current TBPET systems do not provide this information, increasing the PE at large angles~\cite{schmall2016parallax, surti2020total}. 

Although techniques exist to compensate for CS and PE~\cite{zaidi2007scatter}, their overall effect is to degrade the contribution to image reconstruction of gammas emitted at high angles and therefore to partially negate the increase in sensitivity afforded by the longer axial coverage of a TBPET. 
%Notice that TOF-PET scanners also suffer from PE. 

%A possible cost-effective solution is represented by fast plastic scintillators, which are actively being explored for total-body imaging~\cite{moskal2020prospects}. However, their low detection efficiency requires several layers of material and the LOR selection resides mainly on the timing information, given the poor energy resolution~\cite{moskal2014test, niedzwiecki2017j}.

In this paper, we propose a new type of TBPET scanner called CRYSP\footnote{The acronym stands simultaneously for CRYStal PET and for CRYogenic ceSium PET.}. CRYSP is based on large monolithic crystals made of pure cesium iodide (CsI) operating at cryogenic temperatures. 
{The light output of these crystals is very high (about 10$^5$~photons/MeV~\cite{mikhailik2015luminescence,Lewis:2021cjv}), which translates into excellent energy resolution{, below 7\% at 511~keV~\cite{Soleti:2024flw}}. This performance, combined with the monolithic form factor, allows to achieve a millimeter-scale spatial resolution in all three dimensions, minimizing the PE. As we will show in the next sections, these features allow to achieve performances that are comparable to the ones of current state-of-the-art TBPET scanners, at a significantly reduced cost.}

This paper is organized as follows. Section~\ref{sec:csi} discusses the potential of pure, cryogenic CsI as scintillator material for PET, and how the choice of this material leads naturally to monolithic crystals. Section~\ref{sec:crysp} describes the CRYSP baseline design and section~\ref{sec:electronics} focuses on the dedicated electronics. The modelization of the apparatus using the Geant4 simulation framework~\cite{GEANT4:2002zbu} is detailed in section~\ref{sec:mc}.
Section~\ref{sec:vertex} discusses our strategy to reconstruct the gamma impact coordinates using convolutional neural networks (CNNs). The main parameters illustrating CRYSP performance following the NEMA guidelines (e.g., sensitivity, NECR, spatial resolution), {compared with a reference detector based on pixelated LYSO crystals, are discussed throughout sections~\ref{sec:sensitivity} to \ref{sec:pileup}. Section \ref{sec:images} offers an evaluation of the {two apparatus imaging} performances using a modified Jaszczak phantom. A discussion of the implications of our findings on the development of TBPET scanners is included in section~\ref{sec:discussion}. Conclusions are presented in section \ref{sec:conclu}.

\begin{table*}[htbp]
\begin{center}
\caption{Properties of scintillating materials commonly used for PET, compared with scintillators based on cesium iodide. Data for LYSO, BGO and CsI(Tl) have been obtained from vendor~\cite{Luxium} and are intended at room temperature. Values for CsI were measured in ref.~\cite{Soleti:2024flw}. Cost is normalized by one radiation length $X_0$.  The range in prices depends on the quality of the material and the manufacturing process, with pixelated blocks being more expensive that monolithic crystals for a fixed volume.}
\label{tab:crystals}
{\small
\begin{tabular}{ c  c  c  c  c  c  c  c }
\hline
% \textbf{Material} & \textbf{$Z_{eff}$} & \textbf{$X_0$ (cm)} & \textbf{Density (g/cm$^3$)} & \textbf{Light yield ($\gamma$/MeV)} & \textbf{Decay time (ns)} & \textbf{Emission peak (nm)} & \textbf{Cost (\euro/cm$^3$)} \\
% \hline
% LYSO         & 66 & 1.14 & 7.1 & 33,000  & 36  & 420 & 45 \\
% BGO          & 74 & 1.12 & 7.1 & 9,000   & 300 & 480 & 25 \\
% CsI(Tl)      & 54 & 1.86 & 4.5 & 54,000  & 1000 & 560 & 5 \\
% CsI, T=300~K & 54 & 1.86 & 4.5 & 5,000   & 15  & 310 & 5 \\
% CsI, T=100~K & 54 & 1.86 & 4.5 & 100,000 & 800 & 350 & 5 \\

Material & $Z_{eff}$ & $X_0$ (cm) & $\rho$ (g/cm$^3$) & LY (ph./MeV) & $\tau$ (ns) & Peak (nm) & Cost (\euro/cm$^3\cdot X_{0}$) \\
\hline
LYSO         & 66 & 1.14 & 7.4 & 33,000  & 53  & 420 & 50-80 \\
BGO          & 74 & 1.12 & 7.1 & 9,000  & 300 & 480 & 25-35 \\
CsI:Tl      & 54 & 1.86 & 4.5 & 54,000  & 1000 & 560 & 7-10 \\
CsI (300~K) & 54 & 1.86 & 4.5 & 5,000   & 15  & 310 & 7-10 \\
CsI (100~K) & 54 & 1.86 & 4.5 & 100,000 & 800 & 350 & 7-10 \\

\hline
\end{tabular}
}
\end{center}
\end{table*}
\section{Methods and materials}
\subsection{CsI as a scintillator material for PET}\label{sec:csi}
Table~\ref{tab:crystals} shows the properties of cesium iodide compared with other crystals commonly used for PET. 
Pure CsI at ambient temperature has been adopted in particle physics for experiments set in high-rate and high-radiation environments, given its fast decay time and radiation hardness~\cite{Atanov:2016hoz, Doroshenko:2005gd}. However, at a first glance, its low light yield (5~photons/keV), UV emission, average radiation length and density make it a sub-optimal candidate for PET. 

Cooling pure CsI to cryogenic temperatures, approximately 100 K, causes its light emission spectrum to shift towards the near ultraviolet range, moving from around 310~nm to 350~nm~\cite{amsler2002temperature}. This process also increases its light yield by a factor of 20, to about 100~photons/keV~\cite{mikhailik2015luminescence, Lewis:2021cjv, Ding:2023pqe}. Unfortunately, its decay time also increases, approximately by a factor of 50, from 15~ns to 800~ns. In a PET scanner, the coincidence time resolution (CTR) between pairs of crystals is determined by the material's emission time profile and amount of light detected. An approximate figure of merit is given by~\cite{conti2009comparison}:

\begin{equation}
    \mathrm{CTR}\propto\sqrt{\frac{\tau}{N_{\mathrm{p.e.}}}},\label{eq:ctr}
\end{equation}
where $\tau$ is the decay time constant and $N_{\mathrm{p.e.}}$ is the number of detected photoelectrons.

Using eq.~\eqref{eq:ctr} to compare the CTR for CsI operating at ambient temperature and at 100 K we obtain:

\begin{align}\label{eq:ctrx}
    \mathrm{CTR}_{100~\mathrm{K}}  &\approx  \sqrt{\frac{800~\mathrm{ns}}{15~\mathrm{ns}}\cdot\frac{5~ \mathrm{p.e./keV}}{100~\mathrm{p.e./keV}}}\cdot  
    \mathrm{CTR}_{300~\mathrm{K}}  \\&\approx 1.6 \cdot \mathrm{CTR}_{300~\mathrm{K}}.\nonumber
\end{align}
Thus, since the time resolution at room temperature is at the nanosecond level~\cite{kubota1988new}, effective coincidence timing should be possible also at 100 K, in spite of the long decay time, as we will discuss further in section~\ref{sec:eval}.  

In a previous paper \cite{Soleti:2024flw} we measured the energy resolution and coincident time resolution (CTR) for a pair of pure CsI crystals at cryogenic temperatures read out by solid state photosensors (Hamamatsu MPPCs). At $T=104$~K, an energy resolution better than 7\% FWHM and a coincidence time resolution lower than 2~ns FWHM have been achieved. While these are already very promising results, we believe there is still room for improvement, in particular when using large monolithic crystals, SiPMs with higher PDE at 350 nm and optimized electronics. In fact, the intrinsic resolution of cryogenic CsI at 511~keV is approximately 5\%~\cite{moszynski2005energy}, and one of the crystals showed a resolution of 6.3\%, which we will take as default for the studies presented here. For the CTR, we use a conservative estimate of 1.5~ns, based on the 1.84~ns obtained with the tabletop setup { of ref.~\cite{Soleti:2024flw}}.

\subsection{CRYSP}
\label{sec:crysp}
The basic building block of CRYSP is a monolithic crystal, of dimensions $48 \times 48 \times 37~$ mm$^3$, where the crystal thickness corresponds to 2 radiation lengths. All crystal surfaces are polished, and five of the six faces are wrapped in PTFE to ensure a reflectivity $>95\%$. The wrapping adds about 2 mm to the crystal transverse dimensions. Full wrapping with a good reflector is a must for energy resolution, and in the case of CsI, the relatively hard emission wavelength limits the potential reflectors, with PTFE being the best option~\cite{janecek2012reflectivity}. Each crystal is read out by an array of $8 \times 8$ Hamamatsu S14160-6050HS MPPCs~\cite{Hamamatsu}. These photosensors have an active area of $6 \times 6~$ mm$^2$ and a photon detection efficiency of approximately 38\% at the emission wavelength of cryogenic CsI~\cite{Hamamatsu}. Low-temperature SiPM operation is already well established in large-scale particle physics experiments~\cite{Matteucci:2025msy, DUNE:2021hwx, Ieki:2018pbf}. At 77~K, the dark count rate is reduced by several orders of magnitude, afterpulsing and optical cross-talk are suppressed, and the gain remains stable after compensating for the decrease of breakdown voltage with temperature ($\approx 50$~mV/K)~\cite{NepomukOtte:2016ktf, Hamamatsu}. The photon detection efficiency in the blue--UV region is mostly unaffected at cryogenic temperatures, making SiPMs well suited for the scintillation spectrum of CsI~\cite{Iwai:2019scy, Curras-Rivera:2025clp}.

\begin{figure}[htbp]
\centering
\includegraphics[width=0.8\linewidth]{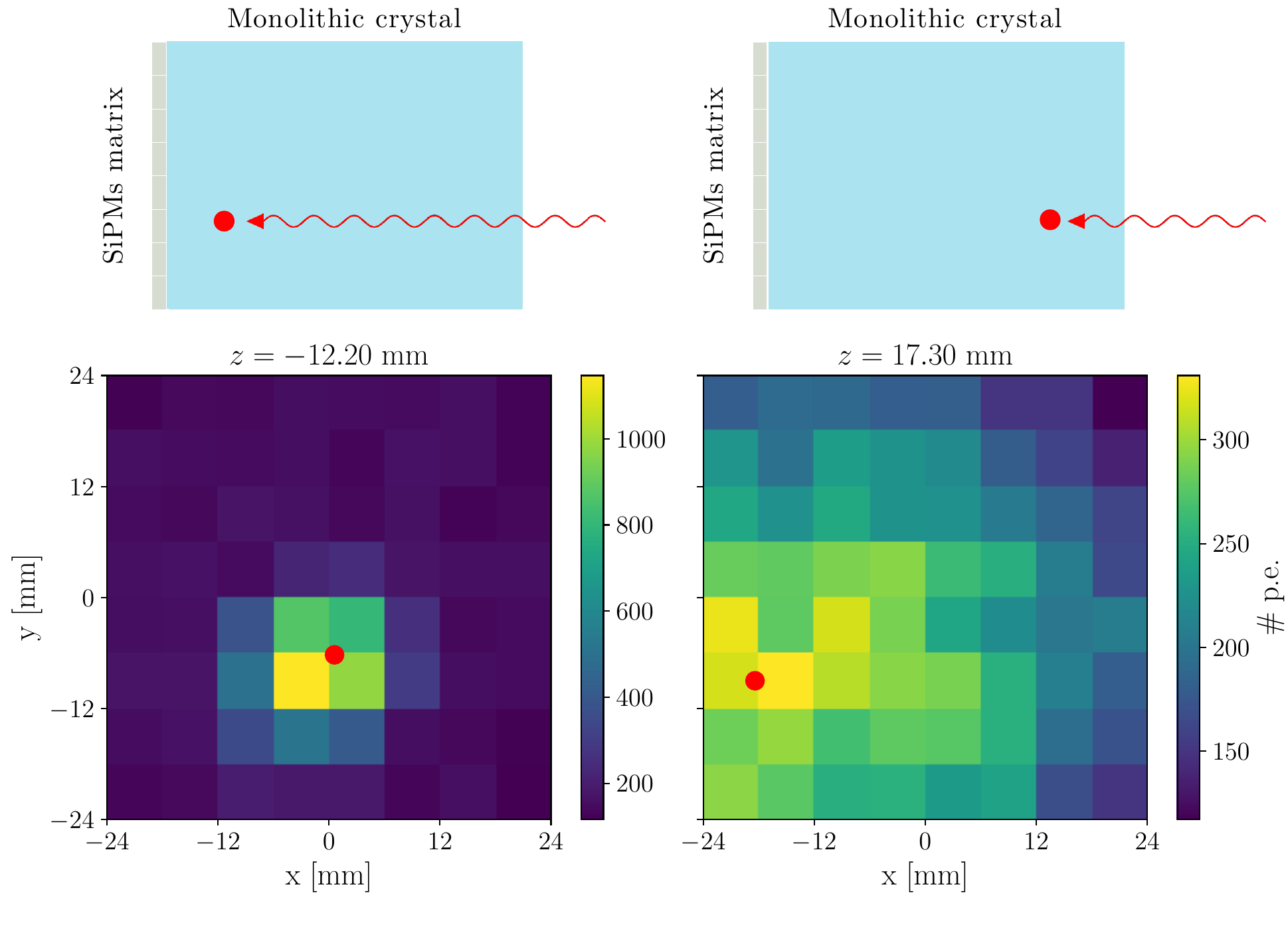}
\caption{Charge collected by the SiPM matrix coupled to a monolithic crystal for two different interactions, one close to the SiPMs plane ($z=-12.20$~mm) and one near the crystal entrance ($z=17.30$~mm). The red dot corresponds to the simulated interaction point. The difference between the two patterns is exploited by a neural network to reconstruct the d.o.i. (here the $z$ coordinate).}\label{fig:image-example}
\end{figure}

Thus, an interaction in the crystal results in an image of $8 \times 8$ pixels, where the intensity in each pixel is determined by the photoelectrons recorded by each SiPM (fig.~\ref{fig:image-example}). 
% Given the high light yield of the scintillator, an option worth considering is using optical concentrators, which funnel the light from a larger entrance into a smaller area. In this way, the photosensors active area  can be smaller, further reducing costs. However, for simplicity, in this baseline proposal we consider photosensors covering the entire exit face of the crystal.

\begin{table}[htbp]
\begin{center}
\caption{CRYSP1M scanner specifications.}
\label{tab:specifications}
\begin{tabular}{ l  c }
\hline
Parameter & Specification  \\
\hline
Crystal size & $48\times48\times37$~mm$^3$\\
SiPM area & $6\times6$~mm$^2$\\
SiPM model & Hamamatsu S14160-6050HS\\
SiPMs per crystal & 64 \\
Detector ring diameter & 77.4~cm\\
PET AFOV & 102.4~cm\\
Detector rings & 20 \\
Crystals per ring & 48\\
Total number of crystals & 960\\
Total number of channels & 61,440\\
Energy resolution & 6.3\% \\
Energy window & 465--555~keV \\
Coincidence time resolution & 1.5~ns\\
Coincidence time window & 4.5~ns\\
Operating temperature & 77~K\\
\hline
\end{tabular}
\end{center}
\end{table}

For definiteness, we consider a CRYSP scanner with dimensions similar to those of the Quadra (CRYSP1M).   
This corresponds to an effective diameter of $\sim 774~$mm and an effective length of $\sim 1024~$mm, which translates to 20 rings, where each ring is comprised of 48 crystals of $50 \times 50 \times 38$~mm$^3$, including wrapping. Thus, CRYSP1M deploys a total of 960 crystals and 61,440 channels (compared with the 243,200 of the Quadra). System specifications are detailed in table~\ref{tab:specifications}. 

{The CRYSP scanner is enclosed within a double‐walled cryostat filled with liquid nitrogen (LN$_2$) that maintains the crystals and SiPMs at approximately 78~K, as shown in figure~\ref{fig:cryspDesign}. The estimated thermal load of 365 W corresponds to an LN$_2$ evaporation rate of 7.6 L/h, resulting in an annual consumption below 70,000 L ($\approx30,000$~€/year). The inner and outer vessels are separated by vacuum layers with multilayer insulation, and all SiPM signals are extracted via compact PCB feedthroughs validated in similar cryogenic environments~\cite{ferrario2022status,PETALO:2024byf}. These mechanical and thermal parameters define the stable operating temperature assumed in simulation and demonstrate the feasibility of continuous cryogenic operation.}

%Figure \ref{fig:cryspDesign} shows a conceptual design. The scanner is hosted in a cryostat. The inner volume of the cryostat is filled with liquid nitrogen that bathes the crystals and the SiPMs, keeping them at approximately 78~K. This implies that dark current in the SiPMs is reduced, and its effect negligible, given the very high yield of the scintillator. The outer volumes of the cryostat are at vacuum and provide thermal isolation. 
%The readout boards are hosted in the outer volume. The liquid nitrogen reservoir which is used to fill the inner part of the cryostat is also used to provide cooling to the readout boards, via cold links. 

% The total estimated heat loss of the cryostat is 20~W. Out of these losses, 70\% are associated with the inner instrumentation of the SIPMs that have a direct thermal path between the inner and outer vessels. The other 30\% are radiation losses assuming a 10 layer Multi-layer Insulation of aluminized Mylar surrounding the inner cryostat. The volume of liquid nitrogen evaporated per hour due to 20 W of heat loss is ~0.45 L/h making a yearly consumption of ~3942 L/year. The typical costs of liquid nitrogen are below 0.6~€/L in facilities with on site liquid nitrogen storage tanks, therefore the associated running costs will be close to 2,000~€/year. 

\begin{figure*}[htbp]
\centering
\includegraphics[width=0.45\linewidth]{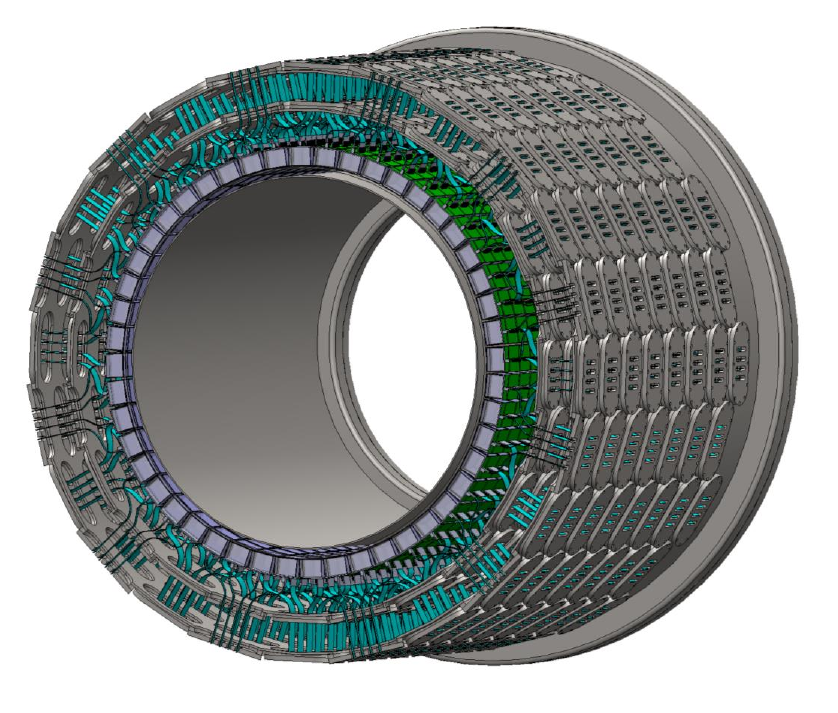}
\includegraphics[width=0.5\linewidth]{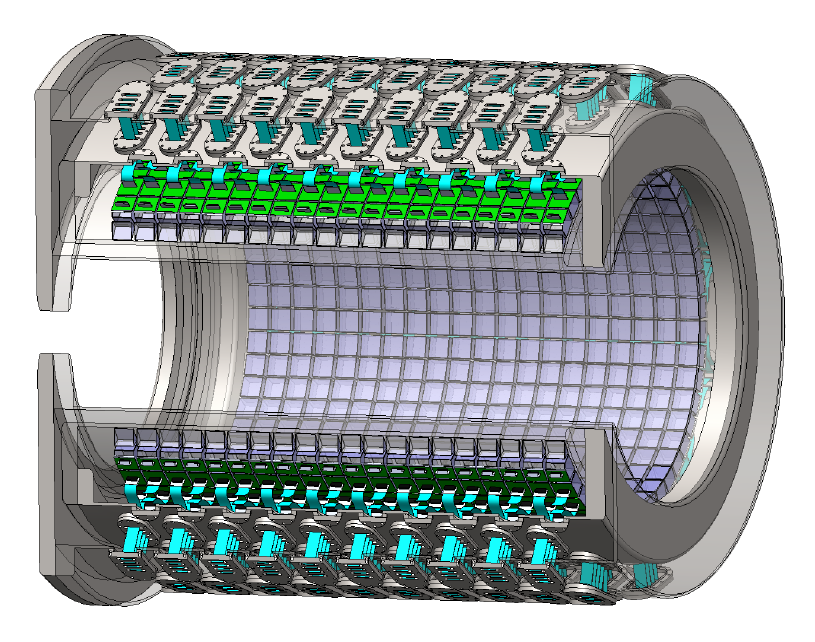}
\caption{CAD drawings of two sections of the CRYSP conceptual design, showing the different parts of the scanner. The crystals, in purple, are placed in a volume filled with liquid nitrogen and read out by a matrix of SiPMs, in green. This volume is separated from the environment by a thin volume at vacuum on the inner side, facing the patient, and a larger volume at vacuum on the outer side, which houses the PCB feedthroughs (in blue). These two vacuum chambers provide thermal insulation to the liquid nitrogen volume.}\label{fig:cryspDesign}
\end{figure*} 
 
% Each SiPM board is connected to a simple feedthrough (e.g., a potted rigid PCB circuit), that allows %passing the signals from the inner to the outer volume of the cryostat. 
% extracting the signal outside the cryostat, where the readout electronics are placed. The connection in between the inner and external feedthrough will be done by flexible cables allowing for the connection and mounting of the different flanges. This arrangement has already been validated in similar environments and does not represent a particular technical challenge~\cite{ferrario2022status}. 
%Each crystal is connected to a readout board which includes an ASIC capable of digitizing the signal of 64 channels as well as providing a time signal (e.g., by taking an OR of the rise time in each channel) and a FPGA which processes the digitized signal. 

When a gamma interacts in a crystal, the readout ASIC digitizes the charge in each of the 64 channels and sends $64 \times 2$~numbers to an FPGA, corresponding to the integrated charge and the timestamp value, as described in section~\ref{sec:electronics}. The FPGA  computes the time of interaction assigned to the crystal (e.g., by taking an OR of the individual timestamps), as well as the total energy and the interaction vertex. The latter is achieved by passing the digitized signals (the $8 \times 8$ pixels comprising the ``image'' provided by the crystal) in real-time through a pre-trained CNN running on the FPGA itself (see section~\ref{sec:vertex}). Thus, the information provided by each crystal is reduced to five numbers: time, energy and the three coordinates of the interaction vertex.

% \begin{figure}[htbp]
% \centering
% \includegraphics[width=0.4\linewidth]{images/ft.png}
% \caption{Detail of the scanner showing the connection between four SiPM boards (in green), coupled to four monolithic crystals (in purple), and the exterior. In this conceptual design, the feedthroughs consist of rigid PCBs acting as feedthroughs connected to flexible cables in the vacuum region to allow for the inner connection the cables and the mounting of the flanges.  A third, thinner vacuum chamber (shown at the bottom in this illustration) separates the liquid nitrogen-filled chamber from the external environment near the patient.}\label{fig:pcbFT}
% \end{figure}

Given the relatively long decay time of CsI ($\tau \approx 1$~\si{\micro\second}), integration of the full charge requires roughly $3 \tau \approx 3~$\si{\micro\second}.  On the other hand, the very high luminosity of the crystal results in a fast rise time, allowing for a CTR in the range of 1-2 ns, as detailed in section~\ref{sec:csi}. Consequently, the system can accommodate high event rates. While this CTR value is too large to support TOF measurements, this limitation has an upside: the requirements on the electronics, which are usually demanding in TOF-PET systems~\cite{schaart2021physics, pourashraf2021investigation}, can be relaxed, further reducing costs.
 
Although the CTR is good enough to resolve coincidences, a fraction of events will still be lost or misreconstructed because of pile-up (e.g., events in which a second gamma interacts in the crystal before the integration of the energy of the first event is over). 
A dedicated pile-up processor is described in section~\ref{sec:pileup_proc} and its effect is quantified in section~\ref{sec:pileup}.

%{A possible improvement of the CTR could be achieved by detecting Cherenkov photons, which are emitted promptly. In this regard, CsI shows promise due to its low self-absorption and relatively high refractive index in the near UV region (1.9 at 350~nm), where Cherenkov emission is more abundant. A preliminary simulation, assuming a single photoelectron time resolution of 100 ps, suggests that a CTR of around 400 ps could be achieved, potentially enabling TOF measurements. However, this estimate require careful experimental verification and, as a conservative assumption, the CTR used in the simulation is set to 1.5~ns.}

\subsection{Electronics for CRYSP}\label{sec:electronics}
The baseline design of CRYSP1M employs front-end integration into an application-specific integrated circuit (ASIC) to bring the information processing closer to the sensors. Therefore, the functionality of the system can be enhanced introducing new features which otherwise could not be implemented offline due to the high amount of data needed to be sent to the data acquisition (DAQ) system. The proposed detector architecture enables a customization of the front-end design that exploits detected signal characteristics allowing for unique features such as event pile-up processing and optimized timestamp generation.

\subsubsection{Timestamp}
Any PET-oriented front-end requires a timestamp generation system with a suitable resolution in order to enable event coincidence filtering. Although CRYSP relies on energy resolution to reject Compton-scattered gammas, an enhanced coincidence filter will help to minimize accidental LORs. High-precision timestamp generating systems depend on accurate clock distribution and synchronization schemes. Moreover, this dedicated hardware must be integrated all along the DAQ data flow, even at the front-end level. This requirement implies a significant increase in costs related to electronic systems. As a consequence, a trade-off between system synchronization complexity and CTR must be observed. In order to improve the latter, a customized time-walk correction mechanism has been proposed for our integrated front-end.

The usual strategy is to approximate timestamp values by $T_0$ crossing time, whose precision is limited by timing resolution. A simple Time-to-Digital Converter (TDC) scheme can easily achieve resolutions below 100~ps with relatively slow clocks (20~ns period)~\cite{kalisz2003review} using affordable ADC converters (less than 9 bits). However, there is a significant effect related to signal slope behavior (\emph{time walk}) that blurs effective timestamp resolution~\cite{time_walk}. In order to reduce this effect a very low threshold can be employed. Therefore, false pulse detection probability would increase due to noise in the input stages of the front-end. Introducing control mechanisms to reject false pulse detection would increase hardware complexity. However, a different approach will be introduced in the development of this integrated front-end: different time values of the pulse waveform will be used to predict an enhanced timestamp value.

The proposed time-walk correction algorithm relies on several threshold crossing times of the rising slope of the pulse signal (see fig. \ref{fig:TW_correction}). The time intervals between these points along with the first (and lowest) threshold crossing point $T_0$, can be used to predict the true origin of the pulse signal:

\begin{equation}\label{eq:1}
    T_S = C_0 \cdot T_0 + C_1\cdot\Delta T_1 + C_2\cdot\Delta T_2 + \dots + C_n\cdot\Delta T_n,
\end{equation}
where $\Delta T_1, \Delta T_2, \dots, \Delta T_n$ correspond to the difference between consecutive threshold crossing times and $T_S$ to the timestamp. The required weights $C_0, C_1, \dots, C_n$ can be estimated through simulation using a simple multiple linear regression learning algorithm.

Furthermore, the final timestamp value can be computed in an analogue fashion, reducing the amount of digital data to a coarse clock period counter and the corrected TDC output value.

This mechanism is implemented on a per-channel basis, meaning the precise timestamp of the event recorded at the detector level is computed off-chip. Typically, the first timestamp among a group of channels is selected as the most accurate event timestamp.

\begin{figure}[htbp]
    \centering
    \includegraphics[width=0.6\linewidth]{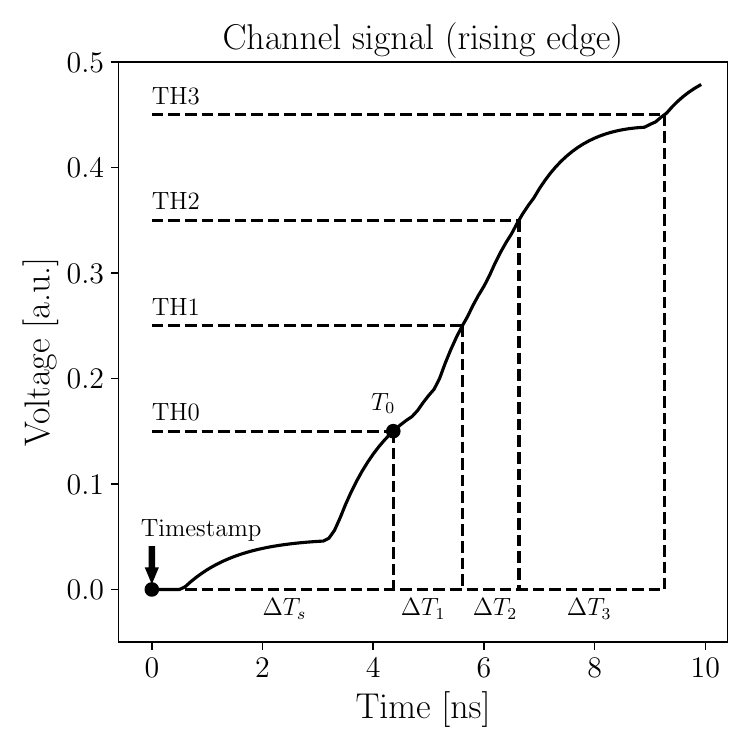}
    \caption{The time-walk correction principle for the rising edge of a signal example. The time intervals between several threshold crossing times and the first (and lowest) threshold crossing point $T_0$ can be used to predict the true origin of the pulse signal.}
    \label{fig:TW_correction}
\end{figure}

\subsubsection{Pile-up processor}\label{sec:pileup_proc}
{Due to the long decay time of cryogenic CsI (approximately 1~\si{\micro\second}) and the use of monolithic crystals, pile-up is a significant concern in CRYSP.}
Thus, a novel pile-up processor based on a peak detection mechanism~\cite{pileup_methods} has been developed for the integrated front-end. For clarity, the method is illustrated using the common scenario of two overlapping pulses in a pile-up event, although it can be extended to handle multiple overlapping pulses.

The peak detection system has been enhanced to extract information from several key signal points, identified by changes in the slope sign of the input signal. To achieve this, an analog derivative computation circuit has been proposed to generate the necessary control signals. The frequency response of this circuit is critical, as it directly impacts the accuracy and limitations of the pile-up processor.

In a two-event pile-up scenario, only three key points are required to determine the energy of both events (see fig.~\ref{fig:two_pileup_event}a). Two of these points, $A$ and $C$, represent the peak positions of the first and second events, respectively. The third point, $B$, corresponds to the start of the second event, which overlaps with the tail of the first event. In an event without pile-up, the peak magnitude is directly related to the event's energy, as the signal shape consistently follows the same time distribution function. Consequently, the energy of the first event can be accurately determined from the amplitude at point $A$.

However, the energy of the second event cannot be directly inferred from the amplitude at point $C$, as the second peak is affected by the downward slope of the first event. To address this, a correction factor must be introduced to accurately determine the true peak amplitude of the second event.

This correction is based on $B$ point time position related to first event starting point, and rise time value of the pulses which is a known constant value. A correction method has been proposed to obtain the right amplitude for the second pulse peak at different pile-up delays using statistical data.

\begin{figure}[htbp]
    \centering
    \subfloat[]{\includegraphics[width=0.57\linewidth]{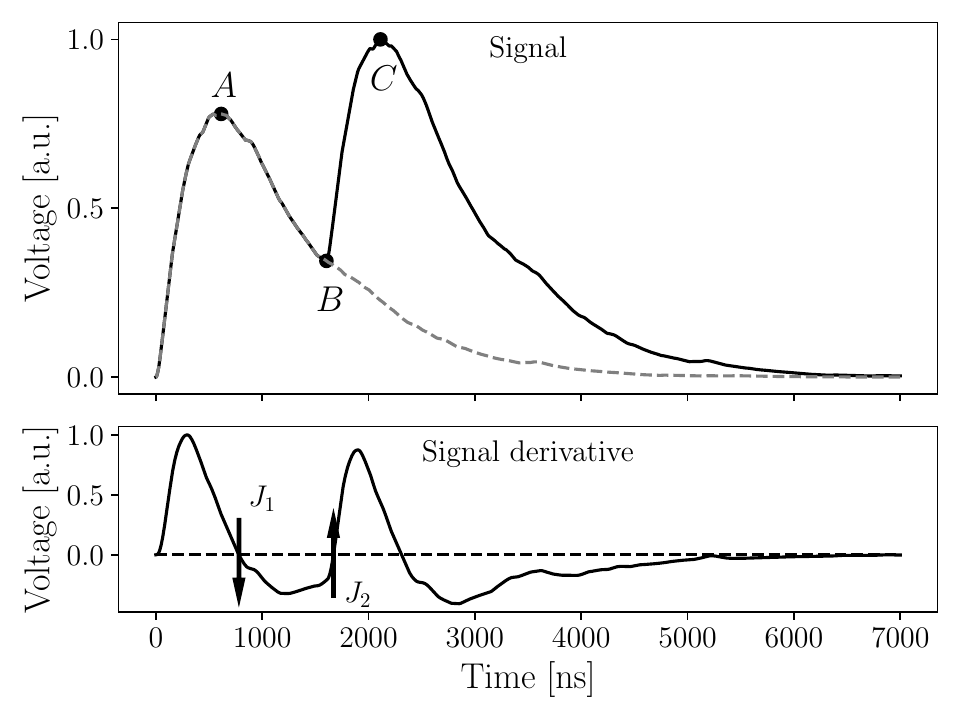}}
    \hfill
    \subfloat[]{\includegraphics[width=0.38\linewidth]{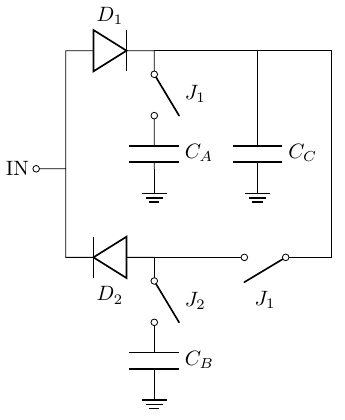}}

    \caption{(a) Example of signal with pile-up (top) and its derivative (bottom), showing the latching times for $J_1$ and $J_2$. (b) Modified multiple peak detector for pile-up processing. The $C_A$ capacitor stores the first peak, latched by $J_1$ upon the derivative signal's positive-to-negative transition. $J_1$ disables $C_B$ charging path, turning it into a falling peak detector. $J_2$, triggered by a negative-to-positive transition, latches $C_B$ to point $B$, while $C_C$ detects the second peak with $D_1$, storing its maximum value.}
    \label{fig:two_pileup_event}
\end{figure}

This operation is an approximation to a deconvolution of the down slope effect of the first pulse, and can be computed offline after data acquisition. Only three data must be converted to digital and sent to the DAQ system: $A$ peak value, $B$ time position and $C-B$ difference (second event peak amplitude to be corrected offline). However, these data must be accurately sampled and latched timely enough to accomplish the ADC operations. The proposed algorithm can be implemented as a modified multiple peak detector (fig.~\ref{fig:two_pileup_event}b) whose control signals rely on the aforementioned analog derivative computation. First of all, the capacitor $C_A$ stores the first peak and the $J_1$ control signal latches its value, which prevents it from losing peak information when the signal rises above it again. $J_1$ will be generated with the first change of the derivative signal from positive-to-negative values. At the same time, $J_1$ breaks $C_B$ charge path through $D_1$, so it turns into a falling peak detector along with $D_2$. When $J_2$ is activated by a negative-to-positive change in the derivative signal, $C_B$ gets latched with $B$ point value.
In the second event, $C_C$ acts now as a peak detector along with $D_1$, then the second peak value gets stored in this capacitor until the maximum value is reached.

Given the timing characteristics of the detector signals and the method used to generate the peak detector control signals, there is a limit to the allowable time overlap between pile-up pulses. A conservative estimate on this limit gives a value of approximately 500~ns. %
{The effect of the unresolved pile-up is quantified in detail in section~\ref{sec:pileup}}.

\subsection{Modeling CRYSP}
\label{sec:mc}
A detailed microphysical simulation of CRYSP has been performed using the Geant4 simulation framework~\cite{GEANT4:2002zbu}. Two different applications have been implemented: one detailing a single monolithic crystal read out by a matrix of $8\times8$ SiPMs of size $6\times6$~mm$^2$ and one with the baseline design of a full CRYSP1M scanner.

The first application was used to study in detail the performances of monolithic crystals. The emission spectrum of CsI at cryogenic temperatures and the photon detection efficiency of the photosensor were {set} according to specifications~\cite{Hamamatsu}. Several possible wrapping materials (e.g., PTFE, ESR~\cite{loignon2017reflectivity}) and coverage combinations were tested. The data obtained with the experimental apparatus of ref.~\cite{Soleti:2024flw} were used to validate the simulation. Optical photon generation was included in the application and the best agreement between data and Monte Carlo was obtained with the \texttt{LUT} optical model~\cite{5485130}. The results of the simulation were then used to develop the vertex reconstruction algorithm described in section~\ref{sec:vertex}. 

The second application includes the full geometry of CRYSP1M and was used for the performance evaluation detailed in section~\ref{sec:eval}. Positrons inside the phantoms were generated with a kinetic energy corresponding to the $^{18}$F decay spectrum. 
However, simulation of optical photons is too computationally intense for a full detector. Thus, the energy deposition in the crystals is smeared with a Gaussian FWHM of 6.3\%, corresponding to the experimental energy resolution obtained in ref.~\cite{Soleti:2024flw}. {The light‐time emission profile and coincidence time resolution (CTR) were applied separately at the analysis stage. A scintillation decay constant of 800~ns and a CTR of 1.5~ns FWHM were adopted, consistent with the experimental results reported in ref.
~\cite{Soleti:2024flw}. The decay time was implemented by time-distributing the detected photoelectrons according to an exponential emission profile with the corresponding decay constant, thereby reproducing the temporal response of the photosensor to the scintillation pulse. Dead-time effects were modeled through the integration and pile-up processing described in section~\ref{sec:pileup_proc}. Interactions occurring within 500~ns were classified as unresolved pile-up and treated as a single event, while pulses separated by more than 500~ns were assumed to be resolvable by the electronics and to preserve the nominal energy resolution.}

{The acquisition parameters used in all performance studies are consistent with those expected for the final detector operation and are summarized in Table~\ref{tab:specifications}. In particular, events were selected within an energy window of 465--555~keV (corresponding to $3\sigma$ around 511~keV) and a coincidence time window of 4.5 ns. These settings were adopted for both sensitivity and imaging simulations to allow a fair comparison across all performance metrics.}

{In order to provide a consistent benchmark, we also simulated a reference LYSO-based PET scanner under the same Monte Carlo framework and acquisition settings used for CRYSP. The geometry and axial coverage were matched to the CRYSP configuration, with detector modules composed of pixelated LYSO crystals ($3\times3\times22$~mm$^3$) read out by SiPMs. The intrinsic parameters were set to 10\% energy resolution, 40~ns decay time, an energy window of 435--585~keV (also here corresponding to $3\sigma$ around 511~keV), and a 350~ps CTR, which lies midway between the performances reported for the Biograph Vision Quadra (230~ps)~\cite{prenosil2022performance} and the uEXPLORER (505~ps)~\cite{badawi2019first}, which enables TOF reconstruction and is consistent with reported performances of state-of-the-art TBPET systems such as the Biograph Vision Quadra~\cite{prenosil2022performance} and uEXPLORER~\cite{badawi2019first}. This virtual LYSO scanner serves as a physically consistent reference for comparing sensitivity, NECR, and spatial resolution results, complementing the image-reconstruction comparison, as shown in section~\ref{sec:eval}.}

% \begin{figure}[htbp]
% \centering
% \includegraphics[width=0.6\linewidth]{images/geant4.png}
% \caption{Geometry simulated with Geant4 of the CRYSP1M detector with a modified Jaszczak phantom placed at the center of the FOV. An electron-positron annihilation inside the phantom produces two gamma rays which are detected by scintillating crystals. Optical photons are drawn here but not simulated in the full CRYSP1M geometry.}\label{fig:geant4}
% \end{figure}

\subsection{Reconstruction of interaction vertex in CRYSP}
\label{sec:vertex}

In a pixelated detector, the reconstruction of the interaction vertex (iV) usually requires a single pixel fired in the detector unit, typically built as an array of several pixels (e.g., in the Quadra scanner, a \emph{mini block} is made of $ 5 \times 5$ pixels, and each pixel has transverse dimensions of $3.2 \times 3.2~$mm$^2$, thus a detector unit is a matrix of 25 pixels with dimensions $16 \times 16 ~$mm$^2$). Imposing that  the energy deposited in the pixel is near 511 keV selects events at or very near the photoelectric peak. 

{The intrinsic spatial resolution in such a configuration is mainly determined by the crystal pitch and thickness. The transverse position of the interaction vertex (iV) can be approximated, in first order, by $\mathrm{FWHM}_t = 2\sqrt{2\ln2}\times t/\sqrt{12}$, where $t$ denotes the transverse dimension of the pixel. For instance, in the Biograph Vision Quadra scanner, $t = 3.2$~mm resulting in $\mathrm{FWHM}_t = 2.1$~mm. Similarly, the d.o.i. resolution can be estimated as $\mathrm{FWHM}_{doi} = 2\sqrt{2\ln2} \times l/\sqrt{12}$, where 
$l$~is the length of the crystal. Considering again the example of the Quadra,  $l=20$~mm yields $\mathrm{FWHM}_{doi} = 13.3$~mm. These expressions represent simple geometrical estimates assuming a uniform interaction distribution and are used as analytical proxies for the intrinsic sampling limits of pixelated PET detectors. }

Thus, while the small size of the pixels guarantees excellent transverse resolution (at the cost of a large number of channels), the d.o.i resolution is typically poor, even for dense crystals, such as LYSO or BGO, since one requires typically a minimum of 2 radiation lengths ($X_0$) for acceptable sensitivity. An additional problem, whose impact becomes increasingly important as the AFOV of the scanner increases, is that of parallax. Gammas entering the detectors at very large angle often do not deposit their energy in the first pixel they encounter on their trajectory, but one or two pixels further. Parallax results in large errors in the reconstruction of the iV. Additionally, gammas can scatter in one pixel, depositing a fraction of their energy, before being fully absorbed in a neighboring pixel. Such events are typically rejected, negatively impacting the sensitivity.

CsI has a longer radiation length ($X_0 \sim18.5$~mm) and smaller photoelectric fraction than the dense, high $Z_{eff}$ scintillators (BGO, LYSO), usually deployed in modern PET scanners. In a pixelated detector this would translate to a larger error in the d.o.i, as well as in a smaller fraction of selected events. 

CRYSP, instead, proposes the use of monolithic crystals, of relatively larger dimensions, read out by an array of large-area SiPMs (see section \ref{sec:crysp}). The simple algorithm used for pixelated systems cannot be applied in this case, since all the 64 SiPMs reading out the crystal have signal when a gamma interacts in the crystal. 
      
The adoption of monolithic crystals for PET and their potential regarding iV reconstruction is not a novel topic~\cite{gonzalez2019novel, gonzalez2021evolution}. However, obtaining a resolution comparable with the transverse dimensions has been historically challenging and classical algorithms fare relatively poorly~\cite{li2008high}. Algorithms based on multilayer perceptrons have also been implemented, providing similar results~\cite{freire2022position, iborra2019ensemble}. Here, instead, we show how excellent resolution in all three coordinates of the iV can be achieved using CNNs, which have been widely adopted for image recognition tasks with extraordinary success~\cite{li2021survey, yamashita2018convolutional}. 

Furthermore, the use of CNNs allows to include in the sample events with a first Compton scatter followed by photoelectric absorption, which deposit the full 511~keV in the same crystal. This is a much larger fraction than pure photoelectric events, up to 60\% of the total events, and is also more significant in CsI than in higher $Z_{eff}$ materials such as LYSO and BGO. Most of these events deposit their energy in two or more clusters in the crystal, but the two most energetic clusters add up to a very large fraction of the total energy deposited, and thus determine the shape of the image. Notably, the first \emph{true} interaction vertex (e.g., the point at which the gamma truly interacts in the crystal) coincides most of the time with the first \emph{apparent} interaction vertex (e.g., the vertex closer to the entry point of the gamma), thus resulting in unbiased, gaussian-distributed residuals between the true interaction vertex of the gamma and the first apparent vertex predicted by our CNN-based reconstruction algorithm, as shown in fig.~\ref{fig:cnn_result}. 

% \begin{figure}[htbp]
% \centering
% \includegraphics[width=0.9\linewidth]{images/cnn_scheme.pdf}
% \caption{Model of the neural network used to predict two iVs in the crystal, taking as input a map of the photoelectrons detected by each SiPM. The convolutional block is repeated three times and then fed to a dense layer, after passing through a flatten and a dropout layer.}\label{fig:cnn_scheme}
% \end{figure}

Our CNN is, therefore, trained to search for two iVs $(x_1, y_1, z_1, x_2, y_2, z_2)$ where the index 1 is assigned to the first apparent interaction vertex. Training the CNN to find two vertices rather than one results in better determination of the position (since, in practice, an algorithm trained to find a single vertex determines the barycenter of the interaction which may be several mm away from the true vertex), and assigning the iV to the first apparent vertex is a good approximation to the true vertex. 
The full network is made of three consecutive convolutional blocks (Conv2D + BatchNorm + LeakyReLU + MaxPooling2D) followed by a Flatten, a Dropout, and a Dense layer yielding 6 outputs $(x_1, y_1, z_1, x_2, y_2, z_2)$.%, as shown in fig.~\ref{fig:cnn_scheme}. 

The network has been trained for 10 epochs on $10^7$ simulated events, with gammas uniformly distributed in the transverse plane. The result is shown in fig.~\ref{fig:cnn_result}, where the difference between the true interaction vertex and the CNN apparent vertex is plotted. The data can be well described by two Gaussians, with an effective resolution of approximately $2\sqrt{2\ln2}\times1.5 \approx 3.5$~mm {\em in all three coordinates}. 

The slightly larger error in the determination of the transverse coordinates, compared with pixelated PET scanners, is compensated by a much smaller error on the d.o.i. Notice, also, that the resolution in the case of a monolithic crystal stays roughly constant with the transaxial angle, since the determination of the iV is not affected by parallax (see section~\ref{sec:spatial_res}). 

For the spatial resolution and image reconstruction studies of CRYSP described in the following sections, the positions of the first interaction of the two gammas within the crystals are smeared with the two Gaussian distributions of fig.~\ref{fig:cnn_result} and are then used directly to define the listmode LORs.  Although a mild depth dependence of the spatial resolution exists (0.2--0.5~mm), in a monolithic detector the d.o.i. is not known \emph{a priori} and must be estimated from the light distribution. Thus, as the reconstruction algorithm does not have access to the true depth, a depth-dependent intrinsic smearing cannot be applied in a realistic system-level simulation. In the pixelated LYSO-based PET system, where no dedicated d.o.i. encoding is assumed, the interaction position is assigned to the geometric position of the crystal.

\begin{figure*}[htbp]
\centering
\includegraphics[width=0.99\linewidth]{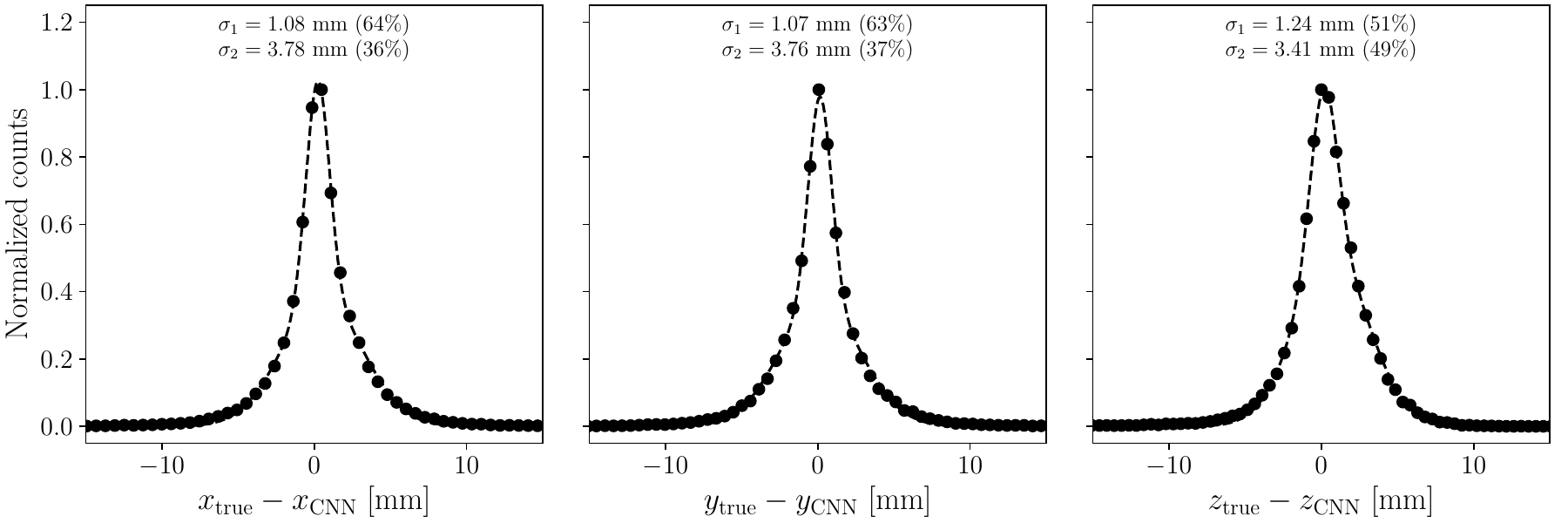}
\caption{Residuals between the true interaction vertex of the gamma and the first apparent vertex predicted by the CNN, in the three axes. The distributions have been fitted with the sum of two Gaussians (dashed line).}\label{fig:cnn_result}
\end{figure*}

\section{Results}\label{sec:eval}
\subsection{Sensitivity}\label{sec:sensitivity}
The performance of CRYSP was benchmarked according to the NEMA NU 2-2018 standard~\cite{national2018nema} using a Monte Carlo simulation of the setup, produced with a dedicated Geant4~\cite{GEANT4:2002zbu} application, as detailed in section~\ref{sec:mc}. 
The sensitivity of CRYSP was measured by simulating a polyethylene phantom with a length of 700 mm, an inner diameter of 2 mm, and an outer diameter of 3 mm. The positron emission was generated uniformly within the inner part of the tubing, which was filled with water. The events were selected in an energy window of 465--555 keV, which corresponds to approximately 3.$\sigma$ around the 511~keV peak, and a coincidence time window of 4.5~ns. These two selection criteria were used also for the rest of the studies of this document, as specified in table~\ref{tab:specifications}. 

\begin{figure}[htbp]
\centering
\includegraphics[width=0.6\linewidth]{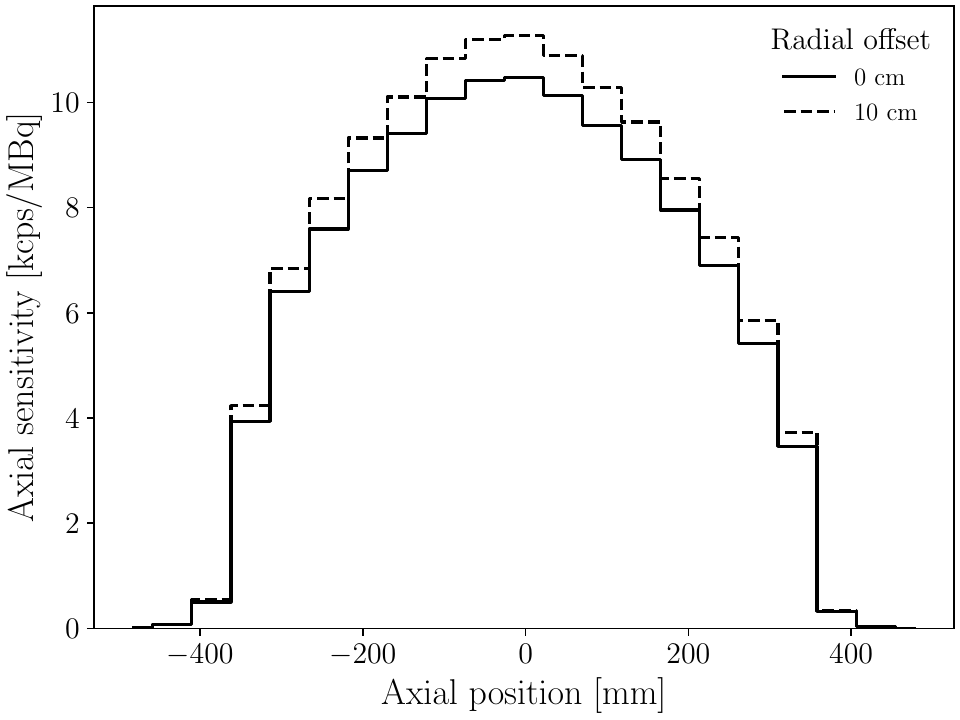}
\caption{Sensitivity as a function of the axial position for the CRYSP scanner with the phantom placed at the center of the FOV (solid line) and with a radial offset of 10~cm (dashed line).}\label{fig:axial_sens}
\end{figure}

According to the NEMA NU 2-2018 protocol, five different simulations were carried out, adding an aluminum sleeve with a thickness of 2.5~mm at each iteration.  Finally, a sensitivity curve was plotted versus the total aluminum thickness to extract the attenuation-corrected sensitivity.

The procedure was repeated in two different configurations, one with the phantom placed at the center of the FOV and one with a radial offset of 10~cm, obtaining $R_{\mathrm{0~cm}}=120\pm5$~kcps/MBq and $R_{\mathrm{10~cm}}=128\pm5$~kcps/MBq, respectively. Although these values are lower than those reported for LYSO-based TBPETs (174 kcps/MBq for the uEXPLORER \cite{spencer2021performance} and 176 kcps/MBq for the Quadra \cite{prenosil2022performance}), the selected events exhibit less Compton contamination due to superior energy resolution{, as detailed in section~\ref{sec:necr}}. The attenuation-corrected sensitivity as a function of the axial position for both radial offsets is reported in fig.~\ref{fig:axial_sens}. This results in reconstructed images of comparable quality at parity of number of events (see section~\ref{sec:images}).  {For reference, our simulated LYSO-based PET scanner yields a sensitivity of $(177 \pm 6)$~kcps/MBq, in excellent agreement with the reported Quadra performance of 175.3 kcps/MBq~\cite{prenosil2022performance}, further supporting the validity of our Monte Carlo model. }

\subsection{Spatial resolution}\label{sec:spatial_res}
Spatial resolution of the scanner has been quantified according to the NEMA standard by simulating a point-like positron source placed at six different positions: in the axial direction at the center of the AFOV and three-eights of the AFOV from the center and in the transverse direction at 1~cm, 10~cm, and 20~cm from the center. The resolution is measured {for both the CRYSP and the reference LYSO-based PET system} in all three directions: radially, axially, and tangentially. Table~\ref{tab:spatial} shows the full width at half-maximum (FWHM) and full width at tenth-maximum (FWTM) amplitudes for the six positions {obtained with the filtered backprojection algorithm. Notably, the CRYSP performance becomes significantly superior to that of the LYSO-based PET at large radial offsets and off-center axial positions, where parallax effects in pixelated detectors degrade spatial resolution.}

\begin{table*}[htbp]
\begin{center}
\caption{FWHM and FWTM in all three directions for point-like sources at six different locations {in CRYSP and in the LYSO-based reference PET, obtained with the filtered backprojection algorithm}.}
\label{tab:spatial}
{\small
\begin{tabular}{llcccccccc}
\hline
\multirow{2}{*}{System} &
\multirow{2}{*}{\shortstack{Axial \\ position (mm)}} &
\multirow{2}{*}{\shortstack{Radial \\ offset (mm)}} &
\multicolumn{3}{c}{FWHM (mm)} &
\multicolumn{3}{c}{FWTM (mm)} \\ \cline{4-9}
& & &                     Radial & Axial & Tangential & Radial & Axial & Tangential \\ \hline
CRYSP & 512 (1/2 of FOV) & 10  & 3.35 & 3.42 & 3.42 & 6.13 & 6.26 & 6.26 \\ 
      & 128 (1/8 of FOV) & 10  & 3.36 & 3.41 & 3.41 & 6.15 & 6.24 & 6.25 \\
LYSO  & 512 (1/2 of FOV) & 10  & 3.52 & 3.61 & 3.61 & 6.44 & 6.60 & 6.60 \\ 
      & 128 (1/8 of FOV) & 10  & 3.51 & 3.62 & 3.61 & 6.39 & 6.60 & 6.58 \\ \hline
CRYSP & 512 (1/2 of FOV) & 100 & 3.39 & 3.43 & 3.52 & 6.21 & 6.28 & 6.44 \\ 
      & 128 (1/8 of FOV) & 100 & 3.43 & 3.42 & 3.52 & 6.28 & 6.26 & 6.44 \\ 
LYSO  & 512 (1/2 of FOV) & 100 & 4.21 & 3.72 & 3.71 & 7.68 & 6.78 & 6.76 \\ 
      & 128 (1/8 of FOV) & 100 & 4.26 & 3.99 & 3.78 & 7.77 & 7.27 & 6.89 \\ \hline
CRYSP & 512 (1/2 of FOV) & 200 & 3.53 & 3.47 & 3.79 & 6.46 & 6.17 & 6.94 \\
      & 128 (1/8 of FOV) & 200 & 3.69 & 3.60 & 3.97 & 6.76 & 6.59 & 7.27 \\
LYSO  & 512 (1/2 of FOV) & 200 & 4.89 & 4.61 & 3.94 & 8.91 & 8.40 & 7.18 \\
      & 128 (1/8 of FOV) & 200 & 5.01 & 4.54 & 3.91 & 9.13 & 8.27 & 7.13 \\ \hline
\end{tabular}
}
\end{center}
\end{table*}

\subsection{Count rates and NECR}\label{sec:necr}

{The count‐rate performance was evaluated for both the CRYSP and the reference LYSO‐based PET scanner using the NEMA NU 2–2018 protocol}. A solid circular cylinder made of polyethylene with an outer diameter of 203 mm, an overall length of 700 mm, and an internal volume of 22,000 mL was simulated. The phantom was placed at the center of the FOV and axially aligned with the scanner. A 6.4 mm hole, positioned 4 mm off the central axis, contained a 700 mm-long tubing (inner/outer diameters of 3.2/4.8 mm) filled with a uniform positron source. In each configuration, $4.5\times10^{6}$ positrons were generated uniformly within the tubing volume, and the acquisition time window was varied to reproduce different activity concentrations between 0 and 40 kBq/mL.

Rates of total, true, scattered, and noise-equivalent counts were calculated according to Section 4 of the NEMA NU 2–2018 standard and are shown in Fig.~\ref{fig:necr}. {A peak NECR of $(1.79\pm0.04)$~Mcps was obtained for CRYSP at an equivalent activity concentration of $14.0$~kBq/mL, while the LYSO-based PET reached $(2.97\pm0.05)$~Mcps at $25.6$~kBq/mL, in good agreement with the reported measurement from Quadra~\cite{prenosil2022performance} (2.96 Mcps for MRD at 27.5 kBq/mL). The corresponding scatter fraction (SF) was found to be approximately 20\% for CRYSP and 40\% for the LYSO-based PET, as shown in Fig.~\ref{fig:necr} (right), indicating that the improved energy resolution of cryogenic CsI effectively reduces scattered coincidences by about a factor of two. Also in this case, the SF for the LYSO-based simulated PET is in good agreement with the Quadra reported measurement~\cite{prenosil2022performance}.}

\begin{figure}[htbp]
\centering
\includegraphics[width=0.48\linewidth]{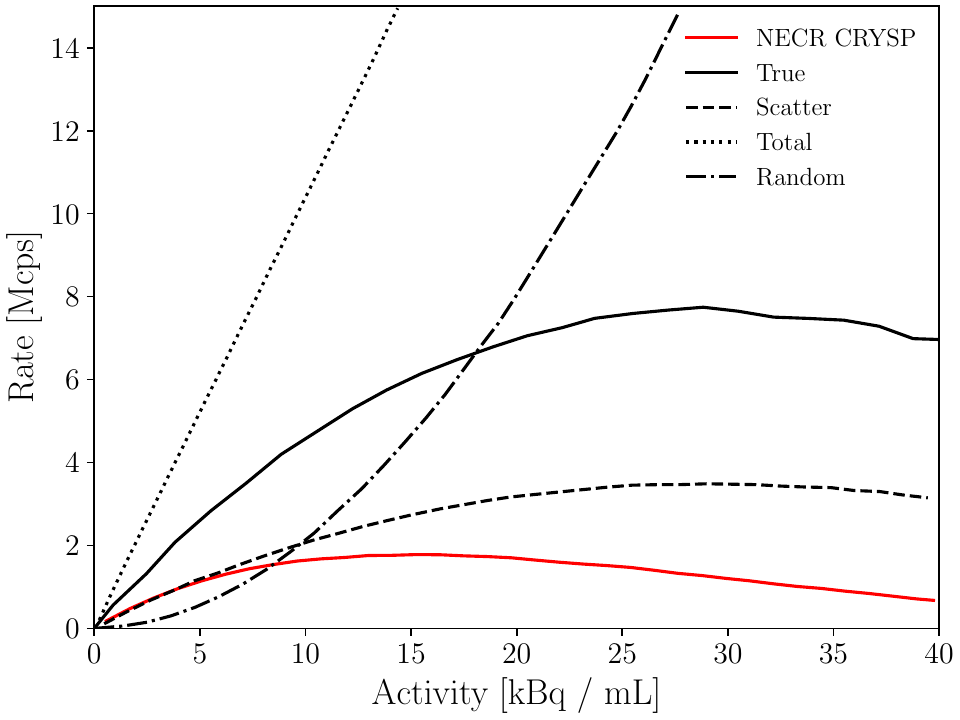}
\quad
\includegraphics[width=0.48\linewidth]{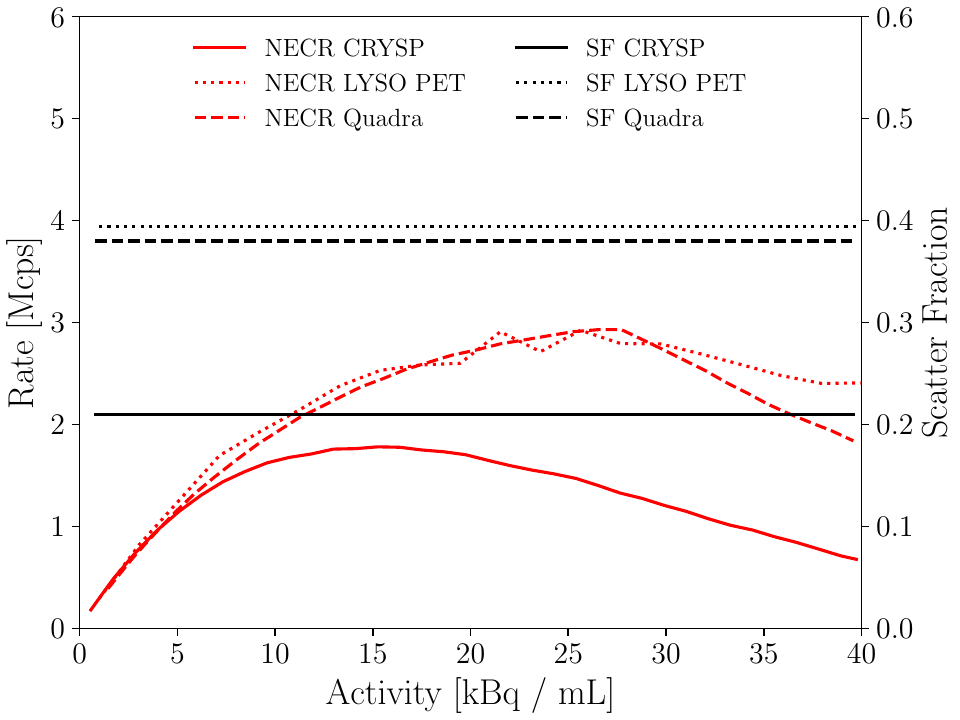}

\caption{Left: true (solid line), scatter (dashed line), random (dash-dotted line), and total (dotted line), and NECR curve (red line) for the CRYSP system. Right: NECR curves (red, left axis) and scatter fractions (black, right axis) for CRYSP (solid), the Biograph Vision Quadra~\cite{prenosil2022performance} (dashed), and a LYSO-based simulated PET scanner (dotted). Both plots are as a function of the activity concentration in the phantom tubing.}\label{fig:necr}
\end{figure}

\subsection{Pile-up}\label{sec:pileup}
In order to quantify the effect of pile-up more realistically, we simulated a simplified MIRD human phantom~\cite{parach2011assessment} inside CRYSP1M, using approximate relative activities in each organ as detailed in ref.~\cite{Dias2022NormalVF}. Two scenarios are being considered: one where the patient's entire upper body is inside the scanner and another where the head is outside the FOV. In our simulation, signals pass through the pile-up processor described in section~\ref{sec:pileup_proc}. An event is classified as pile-up if the processor fails to resolve a second signal arriving within the integration window of the first. Fig.~\ref{fig:pileup} illustrates the fraction of events with unresolved pile-up as a function of the activity administered to the patient. For a typical PET administered activity of 300~MBq, approximately 20\% of events fall into this category. This fraction can be halved by positioning the patient’s head, where a significant portion of the activity is concentrated, outside the FOV.

A key advantage of TBPETs, however, is their ability to maintain image quality with up to an order of magnitude lower administered dose, as demonstrated by uEXPLORER~\cite{badawi2019first}. At an activity level of 30~MBq, this results in an almost negligible fraction of unresolved pile-up, approximately 2\%. {For comparison, in the reference LYSO-based PET scanner, the much shorter scintillation decay time and smaller detector block area result in a negligible pile-up fraction under the same activity conditions.}

\begin{figure}[htbp]
\centering
\includegraphics[width=0.6\linewidth]{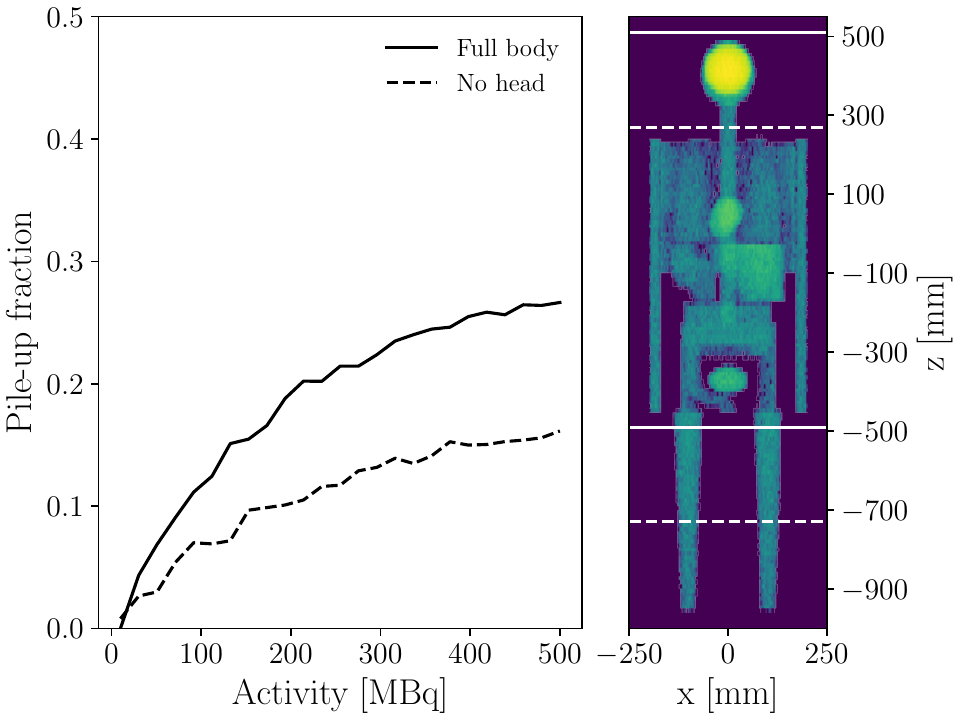}
\caption{Fraction of events with pile-up as a function of administered activity in two situations: one with full upper body inside the scanner (solid lines) and one with the head outside the AFOV (dashed lines). Simulation has been performed with a simplified MIRD model, shown on the right, where the color corresponds to the number of electron-positron annihilations.}\label{fig:pileup}
\end{figure}

\subsection{Image reconstruction in CRYSP}\label{sec:images}

The quality of the images reconstructed by CRYSP represents the chief parameter to be compared with existing solutions, since it encodes sensitivity, spatial resolution, time resolution, energy resolution and NECR performances. To do so, we compare the contrast recovery coefficient (CRC) and signal-to-noise ratio (SNR) of CRYSP with the one of the simulated pixelated LYSO PET, as described in section~\ref{sec:mc}, with and without TOF capabilities.

In order to better assess the impact of Compton scattering with the patient's body, we simulated a modified Jaszczak phantom with an inner diameter of 266~mm (instead of the typical 216~mm) and an axial length of 686~mm (instead of 186~mm), placed at the center of the AFOV. 
{Adopting a phantom with larger length and diameter allows for a more accurate estimation of Compton scattering effects in human patients. This is particularly important in a TBPET, where a significant fraction of detected gammas is emitted at low angles and passes through substantial portions of the body, increasing the likelihood of Compton scattering}. The simulation was performed with $3\times10^9$ electron-positron annihilation events distributed in 10~s of time, corresponding to an activity concentration of 7.9~kBq/mL. This value, although smaller than the peak NECR, has been chosen because it gives a manageable pile-up fraction of less than 20\%. 
%The reconstruction takes into account its effect, but a detailed study of the relation between pile-up, NECR, and image quality is needed in future work. 

\begin{figure}[ht!]
\centering
\includegraphics[width=0.99\linewidth]{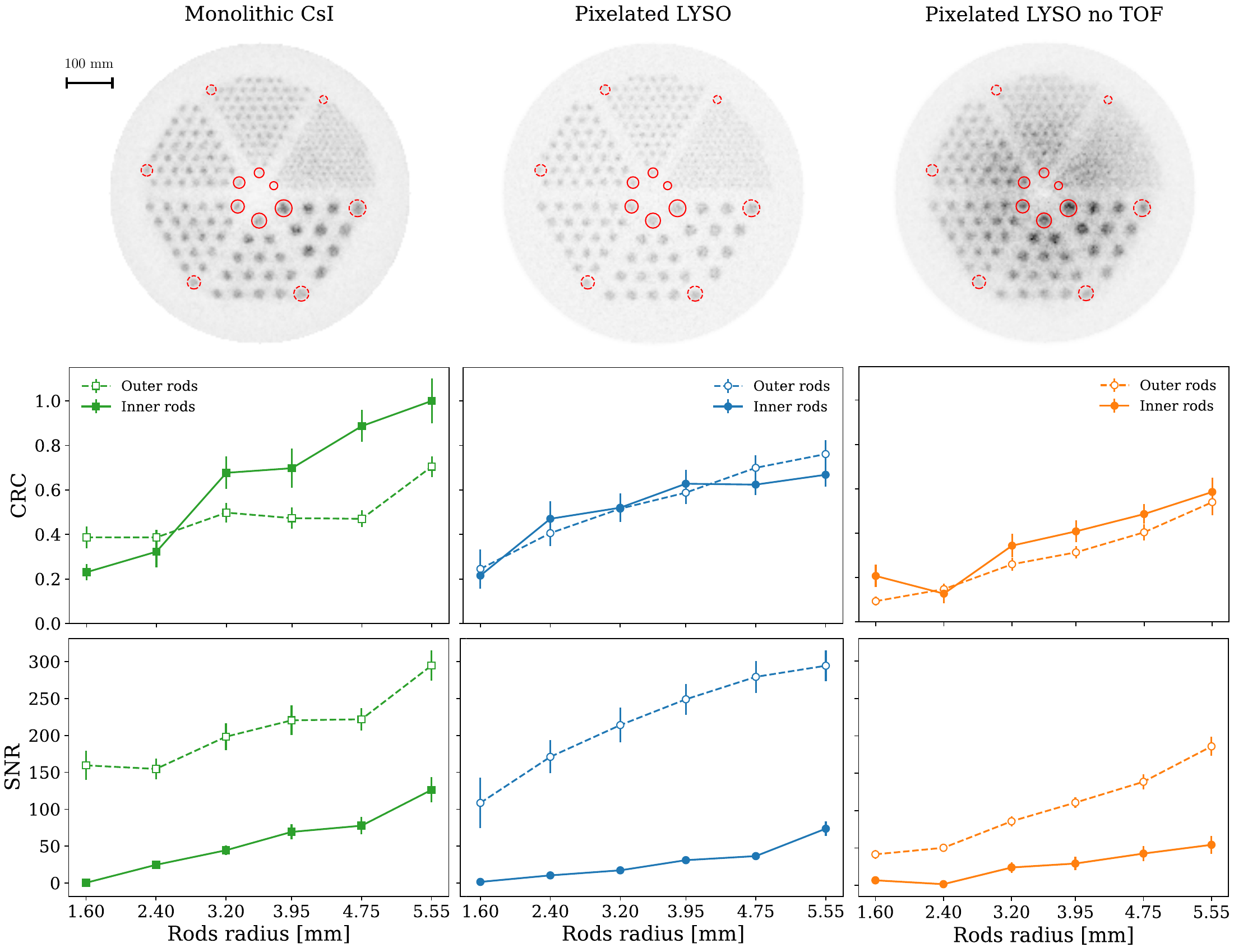}
\caption{Top: reconstructed images with a modified Jaszczak phantom for a PET scanner with monolithic cryogenic CsI crystals (left) and pixelated LYSO crystals with TOF (center) and without TOF (right). Normalization and attenuation corrections are applied and the color scale is the same. Bottom: contrast recovery coefficients and signal-to-noise ratios as a function of the radius of the inner (solid) and outer (dashed) rods. The regions being considered correspond to the red circumferences shown above.}\label{fig:crc}
\end{figure}

The CRC and SNR were calculated for all rod radii, both for rods placed near the center of the phantom (\emph{inner rods}) and for rods placed closer to the surface (\emph{outer rods}), as shown in fig.~\ref{fig:crc}. Reconstruction has been performed with the \texttt{parallelproj} Python package~\cite{schramm2024parallelproj}, which allows to run a highly-parallelized implementation of the MLEM algorithm on GPUs. The reconstruction voxel size was set to 1~mm in all three axes, a value close to the spatial resolution obtained by the CNN, and the algorithm was run for 10 iterations, where the CRC and SNR values plateau. Normalization and attenuation corrections were applied in both cases and no point-spread function modeling was used.  A listmode MLEM algorithm was employed for both the CRYSP and the LYSO-based PET reconstructions. All oblique LORs arising from the 100~cm axial coverage were included without any axial compression or sinogram rebinning. No time-of-flight (TOF) information was used in the CRYSP case, given its CTR of 1.5~ns, which is insufficient to provide a meaningful TOF constraint. For the LYSO-based PET system, reconstructions were performed both with TOF information (assuming a 350~ps CTR) and without TOF, in order to assess the impact of TOF on image quality. Voxel size and projector settings were kept identical across the two cases.

The plots show that a PET scanner based on monolithic CsI crystals has performances that can be compared with ones employing LYSO pixels. Most interestingly, the CRC for rods placed near the center of the phantom is significantly better with CsI: in this case, in fact, the probability for a gamma to undergo Compton scattering in the phantom material is larger, and the significantly better energy resolution of cryogenic CsI {and its improved d.o.i. resolution} more than compensates for the lower sensitivity and lack of TOF. For the inner rods, the CRC goes from 21\% for the smallest radius (1.60~mm) to 96\% for the largest one (5.55~mm). In the case of the outer rods, the variation is smaller, from 39\% to 68\%. As expected, the pixelated LYSO-based PET reconstruction without TOF performs worse than its TOF-enabled counterpart. Moreover, due to the combination of poorer energy resolution and the absence of d.o.i. information, its performance is also inferior to that of CRYSP.

\section{Discussion}\label{sec:discussion}

The development of Total-Body PET (TBPET) apparatus represents a major leap forward in medical imaging. On one hand, it enables true multi-modal and multi-organ imaging; on the other, it allows for significant reductions in radiation dose and/or acquisition time. These improvements open new possibilities, such as routine PET imaging for children, and enable faster patient throughput.

While early demonstrators, such as the uEXPLORER scanner~\cite{cherry2018total}, achieve axial lengths of up to 200 cm, excellent performance can also be obtained with smaller coverage, with 100 cm representing a good compromise for many applications, {as demonstrated by the Biograph Vision Quadra~\cite{prenosil2022performance}.}

However, the high costs associated with TBPETs may limit their widespread adoption. These costs are dominated by the high price of LYSO, the material of choice for virtually all modern PET scanners. LYSO is a rare earth material, and both its production and supply are highly concentrated geographically, raising concerns about future availability and cost.

Another challenge is that increasing the axial length of a PET scanner faces a problem of diminishing returns, linked to two main issues:  
(i) increased multiple scattering as photons travel longer distances through the patient’s body, and  
(ii) the parallax effect, which becomes a major source of reconstruction error for photons traveling at small angles with respect to the $z$-axis.  

The impact of multiple scattering can be mitigated by excellent energy resolution, allowing the selection of photons with energies close to 511 keV, thereby reducing scattering. The parallax effect, in turn, can be {significantly} suppressed by using monolithic crystals combined with modern reconstruction techniques, such as those based on convolutional neural networks (CNNs). Monolithic crystals also offer a substantial cost reduction compared with their pixelated counterparts.

{The balance between d.o.i., TOF, and scatter fraction can be analyzed in terms of their respective contributions to the SNR.
The SNR gain provided by TOF information increases with the patient diameter approximately as $\sqrt{D/\Delta x_{\mathrm{TOF}}}$~\cite{budinger1983time, vandenberghe2016recent}, where $\Delta x_{\mathrm{TOF}} = c\,\Delta t/2$. For a 350~ps CTR ($\Delta x_{\mathrm{TOF}} \approx 5.25$~cm) and a 20~cm NEMA NECR phantom, this corresponds to an SNR improvement of approximately $\times1.95$ relative to a non-TOF system.}
        
Information on the d.o.i.  in CRYSP reduces parallax broadening, resulting in a narrower and more spatially uniform point spread function. While this does not necessarily translate into a global increase in the image SNR, the better spatial resolution can reduce the partial volume effect~\cite{mannheim2023characterization}, thus improving the \emph{local} SNR around small structures by limiting the spread of counts into neighboring voxels. In addition, CRYSP exhibits a substantially lower scatter fraction (approximately 20\% versus 40\% for the LYSO-based configurations), corresponding to an expected SNR benefit of approximately $\sqrt{(1 - 0.2)/(1 - 0.4)} \approx 1.15$. Consistent with these considerations, Fig.~\ref{fig:crc} shows that CRYSP achieves SNR values comparable to those of the pixelated LYSO-based PET across the examined rods, effectively compensating for the absence of TOF and clearly outperforming the non-TOF LYSO configuration.

These considerations point to the possibility of developing a low-cost TBPET system based on an abundant and inexpensive scintillator: CsI. The key realizations are that:  
(a) CsI must operate at cryogenic temperatures, where it offers a light yield three times higher than LYSO and an energy resolution roughly twice as good; and  
(b) the system should use monolithic crystals, which combine lower cost with superior performance at large incidence angles.  

The cost of monolithic CsI can be up to an order of magnitude lower than that of pixelated LYSO, while cryogenic operation (in the well-understood regime of liquid nitrogen) adds no more than about 5\% to the cost of a conventional PET scanner. Importantly, 3D position reconstruction via neural networks in monolithic crystals provides a complete solution to the parallax problem.

On the other hand, CsI exhibits a relatively slow decay time of about 1~\si{\micro\second}. Despite this, its high light yield allows coincidence timing resolutions on the order of the nanosecond, adequate for many PET applications, although still worse than the TOF performance achievable with LYSO. In addition, pile-up can become significant in high-rate environments, but specially designed electronics can largely mitigate this effect, rendering it negligible in medium-rate applications, which are among the strongest use cases for TBPET (e.g., dose reduction).

The results presented here demonstrate that the overall performance of a PET scanner based on monolithic CsI crystals can be competitive with that of much more expensive and complex LYSO-based systems. This offers a viable path to implementing TBPET technology at a cost affordable to most hospitals.

\section{Conclusion}\label{sec:conclu}

This work introduces CRYSP, a novel PET system based on cryogenic pure CsI monolithic crystals, with the potential to achieve high performance at significantly reduced cost. Operating at cryogenic temperatures, CsI provides a light yield of approximately $10^5$ photons/MeV, resulting in an energy resolution better than 7\% FWHM~\cite{Soleti:2024flw}. Despite its relatively slow scintillation decay time, it can achieve a coincidence timing resolution of about 1.5~ns, sufficient for PET applications.

Simulation studies indicate millimeter-scale spatial resolution for interaction vertex reconstruction within a monolithic cryogenic CsI crystal. Overall, the simulated CRYSP1M scanner achieves sensitivity, NECR, and CRCs comparable to existing systems of similar size, despite the absence of TOF capabilities. Notably, CRYSP1M {has the potential to outperform} a pixelated LYSO PET scanner in situations where energy resolution is critical, such as imaging samples with a high fraction of Compton-scattered gamma rays.

While cryogenics is standard in other medical imaging modalities (e.g., MRI), its application in PET devices is new. In contrast to MRI, CRYSP cryogenics requires only a liquid nitrogen bath, enabling the use of conventional, low-cost cooling and signal-extraction techniques (e.g., inexpensive feedthroughs). As a result, cryogenics adds less than 5\% to the overall device cost and can be implemented transparently for the end user.

In summary, CRYSP offers a promising, cost-effective, high-performance alternative for TBPET scanners, with the potential to significantly expand access to advanced PET imaging in both clinical and research contexts.

\ack{The authors are thankful to Georg Schramm for the precious help with \texttt{parallelproj}.}

\funding{SRS acknowledges the support of a fellowship from ``la Caixa Foundation'' (ID 100010434) with code LCF/BQ/PI22/11910019 and of a grant from the Basque Country Government with code PUE24-10. JR acknowledges support from the Generalitat Valenciana of Spain under grant CIDEXG/2023/16.}
% This section is a list of funder names and grant numbers

% \roles{Sample text inserted for demonstration.}
% List author names and the contributions made to the article, using terms from the NISO Contributor Roles Taxonomy (CRediT) https://credit.niso.org

\data{The data that support the findings of this study are available upon reasonable request from the authors.}
% For more information on IOP Publishing's research data policy see: https://publishingsupport.iopscience.iop.org/questions/research-data/

\bibliographystyle{IEEEtran}
\bibliography{biblio.bib}

\end{document}